\documentstyle[12pt,epsfig]{article}
\input tcilatex
\newcommand{\bea}{\begin{eqnarray}}
\newcommand{\eea}{\end{eqnarray}}
\newcommand{\ba}{\begin{array}}
\newcommand{\ea}{\end{array}}
\newcommand{\be}{\begin{equation}}
\newcommand{\ee}{\end{equation}}
\newcommand{\beas}{\begin{eqnarray*}}
\newcommand{\eeas}{\end{eqnarray*}}

\newcommand{\nbox}{{\,\lower0.9pt\vbox{\hrule \hbox{\vrule height 0.2
cm \hskip 0.2 cm \vrule height 0.2 cm}\hrule}\,}}
\DeclareFixedFont{\xiiss}{OT1}{cmss}{m}{n}{12}
\DeclareFixedFont{\ixss}{OT1}{cmss}{m}{n}{9}
\DeclareFixedFont{\cmrnine}{OT1}{cmr}{m}{n}{9}
\newcommand{\CC}{\hbox{\xiiss C\kern-.4emI}}
\newcommand{\RR}{\hbox{\xiiss R\kern-.45emI}}
\newcommand{\ZZ}{\hbox{\xiiss Z\kern-.4emZ}}
\newcommand{\CCs}{\hbox{\ixss C\kern-.4emI}}
\newcommand{\ZZs}{\hbox{\ixss Z\kern-.4emZ}}
\newcommand{\pa}{\partial}

\newcommand{\pasl}{\pa\kern-.55em /}
\def\href#1#2{#2}
\begin{document}

\makeatletter
%\@addtoreset{equation}{section} \renewcommand{\theequation}{\thesection.%
%\arabic{equation}}
\begin{titlepage}

\begin{flushright}
Lab/UFR-HEP / 0204  Revisited\\
\end{flushright}

\vskip 2.5cm

\begin{center} {\Large \bf

NC Calabi-Yau Manifolds in\\ Toric Varieties with NC Torus fibration }
\end{center}

\vspace{1ex}

\begin{center}
{M. Bennai and E.H Saidi \footnote {H-saidi@fsr.ac.ma } \
\\
  [3pt]}

\vspace{5mm} { \it Lab/UFR- High Energy Physics, Physics
Department,\\ Faculty of Science, Avenue Ibn Battouta, PO Box
1014, \\ Rabat, Morocco.}

\end{center}

\vspace{2.5ex}
\medskip
\centerline {\bf Abstract}

\bigskip
\begin{quote}
Using the algebraic geometry method of Berenstein and Leigh ( BL
), hep-th/0009209 and  hep-th/0105229 ), and considering singular
toric varieties ${\cal V}_{d+1}$ with NC irrational torus
fibration, we construct NC extensions ${\cal M}_{d}^{(nc)}$  of
complex d dimension Calabi-Yau (CY) manifolds embedded in ${\cal
V}_{d+1}^{(nc)}$. We give realizations of the NC $\mathbf{C}^{\ast
r}$ toric group, derive the constraint eqs for NC Calabi-Yau (
NCCY ) manifolds ${\cal M}^{nc}_d$ embedded in ${\cal
V}_{d+1}^{nc}$ and work out solutions for their generators. We
study fractional $D$ branes at singularities and show that, due to
the complete reducibility property of $\mathbf{C}^{\ast r}$ group
representations, there is an infinite number of non compact
fractional branes at fixed points of the NC toric group.\\

\bigskip

{\small \bf Key words}: \ \ {\small Toric manifolds and Calabi-Yau
type IIA geometry}, $\mathbf{C}^{\ast r}$ {\small toric group and
Torsion, Non Commutative type IIA Geometry vs} $\mathbf{C}^{\ast
r}$ {\small Torsion, Fractional Branes.}
 \end{quote}
%%%%%%%%%%%%%%%%%%%%%%%%%%%%%%%%%%%%%%%%%%%%%
\bigskip

\end{titlepage}

\newpage \newpage

\section{Introduction}

\qquad Since the original work of Connes et {\it al} on Matrix model
compactification on non commutative (NC) torii \cite{r1}, an increasing
interest has been devoted to the study of NC spaces in connection with
solitons in NC quantum field \cite{r2}, and string field theories \cite
{r1,r2,r3}; in particular in the analysis of $D(p-4)/Dp$ brane systems ($p>3$%
) of superstrings \cite{r6,r7} and in the study of tachyon condensation
using the GMS approach \cite{r8}. However most of the NC spaces considered
in these studies involve mainly NC ${\bf R}_{\theta }^{d}$ \cite{r9}, NC $%
{\bf T}_{\theta }^{d}$ torii \cite{r9,r10}, some cases of ${\bf Z}_{n}$ type
orbifolds of NC torii \cite{r11,r12} and some generalizations to NC higher
dimensional cycles such as the NC Hizerbruch complex surface $F_{0}$ used in
\cite{r13} and some special CY orbifolds. Quite recently efforts have been
devoted to go beyond these particular manifolds by considering particular
examples of CY manifolds ${\cal M}$, especially those given by homogeneous
hypersurfaces with discrete symmetries \cite{r14,r15,r16,r17,r18} in
projective spaces such as the quintic ${\cal Q}$ \cite{r19,r20}. Such
analysis is important for the stringy resolution of singularities; but also
for the study of fractional $D$ branes at these singularities \cite
{r21,r22,r23}. The key point in the construction of NCCY orbifolds by help
of algebraic geometry method, is based on solving non commutativity in terms
of the torsion of discrete isometries of the orbifolds. This idea was
successfully applied in \cite{r15}, for the study ALE spaces and aspects of
the NC quintic; then it has been extended in \cite{r17,r18} for the building
of NC orbifolds ${\cal H}_{d}^{nc}$ of complex $d-$dimension homogeneous
hypersurfaces ${\cal H}_{d}$. In this paper, we want to go one step further
by extending these results to the large class of complex $d$ dimension CY
{\it non homogeneous} manifolds ${\cal M}_{d}$ embedded in singular toric
varieties ${\cal V}_{d+1} $ \cite{r25,r26,r27}. The organization of this
paper is as follows; In section 2, we review general aspects of CY manifolds
${\cal M}_{d}$ embedded in toric varieties ${\cal V}_{d+1}$ and focus on the
study of the type $IIA$ geometry. Sections 3 and 4 are mainly devoted to the
study of the NC type $IIA$ extension of ${\cal M}_{d}$ by introducing \ a NC
toric fibration with ${\cal M}_{d}$ as a base. We first show how NC toric
fibrations may be realised and then derive and solve the constraint eqs for
the NC structure. In section 5, we study fractional branes at singular
points and in section 6, we give our conclusion.

\section{CY Manifolds in Toric Varieties}

We start by recalling that there are different ways for building complex $d$
dimension CY manifolds ${\sl M}$. A tricky way is by embedding ${\sl M}$ in
a toric variety ${\cal V}$; that is a complex\ Kahler manifold with some $%
{\bf C}^{\ast r}$ toric actions. The simplest situation is given by the case
where ${\sl M}$\ is described by complex $d$ dimension hypersurface in a
complex $\left( d+1\right) $ toric variety ${\cal V}_{d+1}$. To write down
algebraic geometry eqs for the CY hypersurfaces which may be singular; one
should specify a set of ingredients namely a local holomorphic coordinates
patch of the toric manifold ${\cal V}_{d+1}$, the group of toric action and
the toric data. To do so, one should moreover distinguish between two kinds
of geometries for the CY manifolds {\sl M}$_{d}$; {\it (1)} the so called
type $IIA$ geometry, to which we will refer to as ${\cal M}_{d} $, and {\it %
(2)} its dual type $IIB$ geometry often denoted as ${\cal W}_{d}$. The
latter is\ obtained from ${\cal M}_{d}$ by exchanging their Kahler and
complex structures following from the Hodge identities $\ h^{1,1}\left(
{\cal M}_{d}\right) =h^{d-1,1}\left( {\cal W}_{d}\right) $ and $%
h^{1,1}\left( {\cal W}_{d}\right) =h^{d-1,1}\left( {\cal M}_{d}\right) $
\cite{r28}. In this study, we will mainly focus on the type $IIA$ geometry.

A nice way to introduce the type $IIA$ geometry ${\cal M}_{d}$ is in terms
of $2D$ $N=2$ field theory which is known to describe the propagation of
type $IIA$ strings on ${\cal M}_{d}$. \ In this approach,\ the ${\cal M}_{d}$
background is constructed in terms of $N=2$\ supersymmetric linear sigma
models involving a superfield system $\left\{ V_{a},X_{i}\right\} $
containing $r$ gauge $N=2$ abelian multiplets $V_{a}\left( \sigma ,\theta ,%
\overline{\theta }\right) $ with gauge group $U(1)^{r}$and $\left(
k+1\right) $ chiral matter superfields $X_{i}\left( \sigma ,\theta ,%
\overline{\theta }\right) $ of bosonic components $x_{i}$. In addition to
the usual kinetic terms and gauge-matter couplings, the linear sigma model
action ${\cal S}\left[ V_{a},X_{i}\right] $ of these fields may have $r$
Fayet Iliopoulos (FI) $D$-terms, which in the language of $N=2$ superfields
read as $\zeta _{a}\int d^{2}\sigma d^{2}\theta d^{2}\overline{\theta }$ $%
V_{a}\left( \sigma ,\theta ,\overline{\theta }\right) $, \ with $\zeta _{a}$
being the FI coupling constants. The superfields action ${\cal S}\left[
V_{a},X_{i}\right] $\ may also have a holomorphic superpotential $%
W(X_{0},...,X_{k})$ given by polynomials in the $X_{i}$'s, which in the
infrared limit, is known to describe the CFT$_{2}$ of string propagation on
the type $IIA$ background \cite{r29}. Let us discuss a little bit this
special geometry.

In the method of toric geometry, where to each complex bosonic field $x_{i}$
it is associated some toric data $\left\{ q_{i}^{a},{\bf \nu }_{i}\right\} $%
, or more generally by taking into account relevant data from the dual
geometry $\left\{ q_{i}^{a},{\bf \nu }_{i};p_{\alpha }^{I},{\bf \nu }%
_{\alpha }^{\ast }\right\} $ \cite{r19,r20,r25}, with ${\bf \nu }_{i}$ and $%
{\bf \nu }_{\alpha }^{\ast }$\ being $\left( d+1\right) $ dimension vectors
of \ ${\bf Z}^{d+1}$ self dual lattice, one can write down the algebraic
geometry equation of the complex $d$ CY manifold ${\cal M}_{d}$. This is
given by a holomorphic polynomial in the $x_{i}$'s with some abelian complex
symmetries. In the simplest situation where the toric manifold is given by
the coset\footnote{%
Toric manifods are generally defined by cosets $\left( {\bf C}%
^{k+1}-U\right) {\bf /C}^{\ast r}$ where \ $U\subset $ ${\bf C}^{k+1}$ is
defined by the ${\bf C}^{\ast r}$\ actions and a chosen a triangulation. In
toric geometry, elements of U are defined by those subsets of vertices,
which do not lie together in a single toric cone.} $\ {\cal V}_{d+1}={\bf C}%
^{k+1}/C^{\ast r}$, $d=k-r$, the complex $d$ \ dimension\ CY hypersurface
reads as,
\begin{equation}
P_{d}\left[ x_{0},...,x_{k}\right] =b_{0}\prod_{i=0}^{k}x_{i}+\sum_{\alpha
}b_{\alpha }\prod_{i=0}^{k}x_{i}^{n_{\alpha i}}.  \label{1}
\end{equation}
where the $b_{\alpha }$'s are complex numbers and where the $n_{\alpha i}$\
powers are some positive integers constrained by the $C^{\ast r}$
invariance. Indeed, under the ${\bf C}^{\ast r}$, $x_{i}\longrightarrow
x_{i}\lambda _{a}^{q_{i}^{a}}$ with $q_{i}^{a}$\ some integers, invariance
of $P_{d}\left[ x_{0},...,x_{k}\right] $ \ requires the $n_{\alpha i}$\
integers to be such that,
\begin{equation}
\sum_{i}q_{i}^{a}n_{\alpha i}=0;\qquad \sum_{i}q_{i}^{a}=0.  \label{2}
\end{equation}
Eqs $\sum_{i}q_{i}^{a}=0$, which ensures the vanishing of the first Chern
class of ${\cal M}_{d}$,\ follow from the ${\bf C}^{\ast r}$ symmetry of the
$\prod_{i=0}^{k}x_{i}$\ monomial while $\sum_{i}q_{i}^{a}n_{\alpha i}=0$
come from invariance of $\prod_{i=0}^{k}x_{i}^{n_{\alpha i}}$ monomials.
Setting $u_{\alpha }=\prod_{i=0}^{k}x_{i}^{n_{\alpha i}}$, eq (\ref{1}) can
be rewritten as $P_{d}\left[ u_{\alpha }\right] =\sum_{\alpha }b_{\alpha
}u_{\alpha }$, where the $u_{\alpha }$'s\ are the effective local
coordinates of the coset space ${\bf C}^{k+1}/{\bf C}^{\ast r}$. As the $%
u_{\alpha }$'s are given by $u_{\alpha }=\prod_{i=0}^{k}x_{i}^{n_{\alpha i}}$%
, it may happen that not all of the $u_{\alpha }$'s are independent
variables; some of them, say $u_{\alpha _{I}}$ for $I=1,...r^{\ast }$, are
expressed in terms of the other $u_{\alpha _{J}}$ variables with $J\neq I$.
In other words; one may have relations type $\prod_{\alpha }u_{\alpha
}^{p_{\alpha }^{I}}=1$, where $p_{\alpha }^{I}$\ are some integers.
Substituting $u_{\alpha }=\prod_{i=0}^{k}x_{i}^{n_{\alpha i}}$ back into $%
\prod_{\alpha }u_{\alpha }^{p_{\alpha }^{I}}=1$, we discover extra
constraint eqs on the $n_{\alpha i}$ and $p_{\alpha }^{I}$\ integers namely,
\begin{equation}
\sum_{\alpha }p_{\alpha }^{I}n_{\alpha i}=0.  \label{3}
\end{equation}
In toric geometry, the $n_{\alpha i}$\ integers are realized as $n_{\alpha
i}=<{\bf \nu }_{i},{\bf \nu }_{\alpha }^{\ast }>=\sum_{A}\nu _{i}^{A}\nu
_{\alpha A}^{\ast }$. In this representation, eqs(\ref{2},\ref{3}) are
automatically solved by requiring the toric data of the CY manifold to be
such that,
\begin{equation}
\sum_{i}q_{i}^{a}{\bf \nu }_{i}=0;\qquad \sum_{\alpha }p_{\alpha }^{I}{\bf %
\nu }_{\alpha }^{\ast }=0.  \label{4}
\end{equation}
Let us illustrate these relations for the case of the assymtotically local
euclidean ( ALE ) space with $A_{n-1}$ singularity. This is a complex two
dimension ${\bf C}^{n+1}/{\bf C}^{\ast \left( n-1\right) }$ toric variety
with a $su(n)$ singularity $u_{0}^{-n}u_{1}u_{2}=1$\ at the origin. From
this relation, one sees that there are three invariant variables $u_{0}$, $%
u_{1}$ and $u_{2}$; but only two of them are independent. Since $r^{\ast }=1$%
, there is only one $p_{\alpha }^{I}$\ vector of entries $p_{\alpha }=\left(
-n,1,1\right) $ and three ${\bf \nu }_{\alpha }^{\ast }$ vectors given by $%
{\bf \nu }_{0}^{\ast }=\left( 1,0\right) $, ${\bf \nu }_{1}^{\ast }=\left(
n,-1\right) $ and ${\bf \nu }_{2}^{\ast }=\left( 0,1\right) $. More
generally, we have the following cases: {\it (a)} No relation such as $%
\sum_{\alpha }p_{\alpha }^{I}{\bf \nu }_{\alpha }^{\ast }=0$ exist,\ that is
all the $u_{\alpha }$'s independent; the algebraic eq of the hypersurface $%
{\cal M}_{d}$ is non singular and reads as $\sum_{\alpha =0}^{d}b_{\alpha
}u_{\alpha }=0$ \ with $d=\left( k-r\right) $. {\it (b)} Generic cases where
there exists $r^{\ast }$\ constraint eqs of type $\sum_{\alpha }p_{\alpha
}^{I}{\bf \nu }_{\alpha }^{\ast }=0$, the $\left( d+r^{\ast }+1\right) $ \
complex variables$\ u_{\alpha }$ are not all of them independent; So the
algebraic geometry eqs defining the singular hypersurface embedded in ${\cal %
V}_{d+1}$ is
\begin{equation}
\sum_{\alpha =0}^{d+r^{\ast }}b_{\alpha }u_{\alpha }=0;\qquad \prod_{\alpha
=0}^{d+r^{\ast }}u_{\alpha }^{p_{\alpha }^{I}}=1;\qquad I=1,...,r^{\ast },
\label{5}
\end{equation}
where $p_{\alpha }^{I}$\ are the integers in eqs(\ref{4}). Note that this
hypersurface has singularities at the $u_{\alpha }=0$; but the introduction
of NC geometry lifts this degeneracy. In the next section, we will study the
NC extension of this geometry.

\section{NC Type $IIA$ Geometry vs NC Torus Fibration}

To start note that NCCY manifolds as built in \cite{r15} may be viewed as a
NC torus fibration with a CY base. Since NC torii have two kinds of
realizations namely rational and irrational representations, we have to
distinguish two cases of NC torus fibrations. The first kind of these
manifolds involve fuzy torii and is exactly the solution obtained in \cite
{r15}. The second class of NC varieties, which extend naturally the previous
one, is the type of solution we want to present here. Before that we will
first recall the BL algebraic geometry method; then we present our solution
by direct use of NC fibration ideas.

\subsection{Rational torus}

From the algebraic geometry point of view, the NC extension ${\cal M}%
_{d}^{nc}$ of the CY manifold\ ${\cal M}_{d}$, embedded in ${\cal V}_{d+1}$%
,\ is covered by a finite set of holomorphic operator coordinate patches $%
{\cal O}_{(\alpha )}=\{Z_{i}^{(\alpha )};1\leq i\leq k\;\alpha =1,2,\ldots
\} $ \ and holomorphic transition functions mapping \ ${\cal O}_{(\alpha )}$%
\ \ to \ ${\cal O}_{(\beta )}$; $\phi _{(\alpha ,\beta )}$: \ ${\cal O}%
_{(\alpha )}\ \ \rightarrow \ {\cal O}_{(\beta )}$. \ This is equivalent to
say that ${\cal M}_{d}^{nc}$ is covered by a collection of NC local algebras
${\cal M}_{d}^{nc}{\cal {_{(\alpha )}}}$ generated by the analytic
coordinate of the ${\cal O}_{(\alpha )}$ patches of ${\cal M}_{d}^{nc}$,
together with analytic maps $\phi _{(\alpha ,\beta )}$ on how to glue ${\cal %
M}_{d(\alpha )}^{nc}$ and ${\cal M}_{d}^{nc}{\cal {_{(\beta )}}}$. The $%
{\cal M}_{d}^{nc}{\cal {_{(\alpha )}}}$ algebras have centers ${\cal Z}%
_{\left( \alpha \right) }={\cal Z}\left( {\cal M}_{d(\alpha )}^{nc}\right) $%
; \ when glued together give precisely the commutative manifold ${\cal M}%
_{d} $. In this way, a singularity of ${\cal M}_{d}{\cal \simeq {Z}}\left(
{\cal M}_{d}^{nc}\right) $ can be made smooth in the non commutative space $%
{\cal M}_{d}^{nc}$ \cite{r24,r18}. This idea was successfully used in the
building of NC extensions of the CY homogeneous hypersurface $%
z_{1}^{d+2}+z_{2}^{d+2}+z_{3}^{d+2}+z_{4}^{d+2}+z_{5}^{d+2}+a_{0}\prod_{\mu
=1}^{d+2}z_{\mu }=0$ \ having ${\bf Z}_{d+2}^{d}$ \ discrete\ symmetries
acting as $z_{i}\longrightarrow z_{i}$.$\omega ^{q_{i}^{a}}$, where the $%
q_{i}^{a}$\ integers satisfy the\ CY \ condition $%
\sum_{i=1}^{d+2}q_{i}^{a}=0,$\ $a=1,...,d$; \cite{r15,r16}. The NC ${\cal H}%
_{d}^{nc}$ extending ${\cal H}_{d}$\ were shown to be given, in the
coordinate patch $Z_{d+2}\sim I_{id}$, \ by the following NC algebra,
\begin{eqnarray}
Z_{i}Z_{j} &=&\theta _{ij}\text{ }Z_{j}Z_{i};\qquad \ i,j=1,...,(d+1),
\nonumber \\
Z_{i}Z_{d+2} &=&\text{ }Z_{d+2}Z_{i};\qquad \ i=1,...,(d+1).
\end{eqnarray}
The $\theta _{ij}$ parameters\ are solved by discrete torsion as $\theta
_{ij}=\omega _{ij}$ $\varpi _{ji}$ with $\varpi _{kl}$ the complex conjugate
of $\omega _{kl}$. The $\omega _{ij}$'s are realized in terms of the $%
q_{i}^{a}$ CY charges and the ${\bf Z}_{d+2}^{d}$ discrete group elements $%
\omega =\exp i\frac{2\pi }{d+2}$ as $\omega _{ij}=\exp i\left( \frac{2\pi }{%
d+2}m_{ab}q_{i}^{a}q_{j}^{b}\right) =\omega ^{m_{ab}q_{i}^{a}q_{j}^{b}}$
with $m_{ab}$ integers. The $Z_{i}$'s form a regular representation of the
discrete group and are solved as
\begin{equation}
Z_{i}=x_{i}\prod_{a,b=1}^{d}{\bf P}^{m_{ab}q_{i}^{a}}{\bf Q}_{a}^{q_{i}^{b}},
\label{e0}
\end{equation}
where ${\bf P=}diag\left( 1,\omega ,...,\omega ^{d+1}\right) $ and ${\bf Q}$
satisfy,
\begin{eqnarray}
{\bf P}^{d+2} &=&{\bf Q}^{d+2}=I_{id},  \label{e01} \\
{\bf PQ} &=&{\bf \omega QP;\quad }\omega =\exp i\frac{2\pi }{d+2},
\label{e02}
\end{eqnarray}
\ and generate the NC rational torus. From this solution, one sees that the
BL manifold ${\cal H}_{d}^{nc}$ is just a kind of a NC torus fibration based
on the commutative CY hypersurface ${\cal H}_{d}$.

\subsection{Irrational torus}

To build the NC type $IIA$ geometry extending eq(\ref{1}), we will mainly
adopt the same method as in \cite{r15,r17} and look for realizations of the
NC variables $Z_{i}$ as
\begin{equation}
Z_{i}\sim x_{i}\prod_{a=1}^{r}U_{a}^{r_{i}^{a}}V_{a}^{s_{i}^{a}}.  \label{e1}
\end{equation}
where $r_{i}^{a}$ and $s_{i}^{a}$\ are positive integers to be determined
later. In this fibration, $U_{a}$ and $V_{b}$ are the generators of a
complex $r$ dimension NC torus satisfying, amongst others, the familiar
relation
\begin{equation}
U_{a}V_{b}=\gamma _{ab}V_{b}U_{a}  \label{e2}
\end{equation}
with $\gamma _{ab}$ some given irrational $C$-numbers. One of the main
differences between this fibration and the rational torus one of \cite{r15};
see also eqs(\ref{e0},\ref{e01},\ref{e02}), is that here $\gamma _{ab}$ is
no longer a root of unity and so the $Z_{i}$'s have infinite dimensional
representations which make such kind of NC extension very special as we will
see when we discuss fractional D branes. For the moment let us expose our
result by first giving the constraint eqs, their solutions and the $Z_{i}$'s
regular representations.

\begin{itemize}
\item  {\sl Constraint Eqs}
\end{itemize}

Extending naively the algebraic geometry method used for ${\cal H}_{d}^{nc}$
to our present case by associating to each $x_{i}$\ variable the operator $%
Z_{i}$, then taking $q_{k}^{a}=0$ and working in the coordinate patch $%
Z_{k}=I_{id}$, the NC type $IIA$ geometry ${\cal M}_{d}^{nc}$ may be defined
as,
\begin{eqnarray}
Z_{i}Z_{j} &=&\theta _{ij}Z_{j}Z_{i},\quad i,j=0,...,k  \nonumber \\
Z_{i}Z_{k} &=&Z_{k}Z_{i}.  \label{7}
\end{eqnarray}
Since ${\cal M}_{d}$\ should be in the centre of ${\cal M}_{d}^{nc}$, it
follows that the $Z_{i}$ generators and the $\theta _{ij}$ parameters should
satisfy the constraint eqs,
\begin{equation}
\left[ Z_{i},\prod_{j=0}^{k}Z_{j}^{n_{\alpha j}}\right] =0,\qquad
\prod_{j=0}^{k}\theta _{ij}^{n_{\alpha j}}=1,\quad \forall i,\qquad \theta
_{ij}\theta _{ji}=1.  \label{8}
\end{equation}
Actually these relations constitute the defining conditions of NC type $IIA$
geometry ${\cal M}_{d}^{nc}$. While the constraint relation $\theta
_{ij}\theta _{ji}=1$\ shows that $\theta _{ij}=\theta _{ji}^{-1}$, non
trivial solution of the constraint eqs $\prod_{j=0}^{k}\theta _{ij}=1$ are
expressed in terms of $\left\{ q_{i}^{a};\nu _{i}^{A};p_{\alpha }^{I};\nu
_{\alpha A}^{\ast }\right\} $data.

\begin{itemize}
\item  {\sl Solving the Constraint Eqs}
\end{itemize}

First of all note that since the $\theta _{ij}$\ 's\ are non zero
parameters, one may set
\begin{equation}
\theta _{ij}=\prod_{a,b=1}^{r}\eta _{ab}^{J_{ij}^{ab}};\quad \eta _{ab}=\exp
\left( \beta _{a}\beta _{b}\right) ;\quad \beta _{a}\in {\bf C,}  \label{9}
\end{equation}
and solve the constraint eqs(\ref{8}) by introducing torsion for the ${\bf C}%
^{\ast r}$ toric actions. Putting these parameterizations back into eqs(\ref{8}%
), one gets the following constraint on the $J_{ij}^{ab}$'s,
\begin{equation}
\sum_{i=0}^{k}J_{ij}^{ab}=0;\quad J_{ij}^{ab}=-J_{ji}^{ab}.  \label{10}
\end{equation}
Moreover as we are looking for solutions to the $Z_{i}$\ operators as in eq(%
\ref{e1}), let us explore what one may call NC ${\bf C}^{\ast r}${\it \ }%
toric actions.

{\it NC }${\bf C}^{\ast r}${\it \ Toric group: }\qquad Recall the ${\bf C}%
^{\ast r}$\ toric group as used in toric geometry is a complex abelian group
which reduce to $U\left( 1\right) ^{r}$ once the group elements $\exp i\psi
_{a}T_{a}$, with parameters $\psi _{a}=\alpha _{a}-i\rho _{a}\in {\bf C}$,
are chosen as $\exp i\alpha _{a}Q_{a}$ where the $\alpha _{a}$'s are now
real numbers. To have NC toric actions, there are two ways to do: {\it (1)}
either through the use quantum symmetries as in case of discrete torsion or
{\it (2)} by introducing CDS non commutative torii \cite{r1}. Let us comment
briefly these two realizations.

1. {\it Quantum \ toric symmetry: }\qquad To illustrate the idea, we
consider the simple case $r=1$. Since ${\bf C}^{\ast }$ is an abelian
continuous group and its representations have very special features, we have
to distinguish the usual cases; (a) the discrete infinite dimensional
spectrum and (b) the continuous one. Both of these realizations are
important for the present study and should be thought of as extensions of
the rational and irrational representations of NC real torii \cite{r10,r12}.

a) {\it Discrete Spectrum}:\qquad Let ${\cal R}_{dis}\left( C^{\ast }\right)
=\left\{ U=\exp i\psi T\right\} $ denote a representation of $C^{\ast }$ on
an infinite dimensional space ${\bf E}_{dis}$ with a discrete spectrum
generated by the orthonormal vector basis $\left\{ |n>;n=\left(
n_{1},n_{2}\right) \in {\bf Z\times Z\sim Z}^{2}\right\} $\footnote{%
We will use the convention notation $n\equiv n_{1}+in_{2}\in {\bf Z+iZ\sim }%
\left( n_{1},n_{2}\right) \in {\bf Z}^{2}$; $|n>$ should be then thought of
as $|n_{1}>\otimes |n_{2}>$\ .}. Here $\psi $ is a complex parameter, $\psi
\in {\bf C}$; and $T$ is the complex generator of ${\bf C}^{\ast }$; the $%
\psi T$ combination may be split as $\psi T=\rho K+i\alpha Q$, where $K$\ is
the generator of dilatations and $Q$ is the generator of phases. For $\psi
=\alpha $ real, ${\cal R}_{dis}\left( C^{\ast }\right) $ reduces to ${\cal R}%
_{dis}\left( U\left( 1\right) \right) =\left\{ U=\exp i\alpha Q\right\} $.
The generator $T$ of ${\cal R}_{dis}\left( C^{\ast }\right) $\ acts
diagonally on the vector basis $\left\{ |n>\right\} $; i.e \ $T|n>=n|n>$\
and so $U|n>=\left( \exp i\psi n\right) |n>$ or equivalently $U=\sum_{n}$ $%
\chi _{n}\left( \psi \right) $ $\pi _{n}$, with characters $\chi _{n}\left(
\psi \right) =\exp \left( i\psi n\right) $. and $\pi _{n}=|n><n|$, the\
projectors on $|n>$ $\pi _{n}$. Since ${\cal R}_{dis}\left( C^{\ast }\right)
$ is completely reducible, its $I_{id}$ identity operator may be decomposed
in a series of $\pi _{n}$ as,
\begin{equation}
I_{id}=\sum_{n}\pi _{n}.  \label{11}
\end{equation}
Like for $U\left( 1\right) $ and ${\bf Z}_{N}$ subgroups, ${\bf C}^{\ast }$
has also a complex shift operator $V_{\tau }$\ acting on $\left\{ |n>;n\in
{\bf Z+iZ}\right\} $ as an automorphism exchanging the characters $\chi
_{n}\left( \psi \right) $. This translation operator which operate as $%
V_{\tau }|n>=|n+\tau >$; with $\tau =1+i$, may also be defined by help of
the ${\bf a}_{\left( n_{1},n_{2}\right) }^{+}=|\left( n_{1}+1\right)
+in_{2}><n_{1}+in_{2}|$ \ and \ ${\bf b}_{\left( n_{1},n_{2}\right)
}^{+}=|n_{1}+i\left( n_{2}+1\right) ><n_{1}+in_{2}|$\ \ step operators\ as $%
V_{\tau }=V_{1}\otimes V_{i}$ with,
\begin{equation}
V_{1}=\sum_{n_{1},n_{2}\in {\bf Z}}{\bf a}_{\left( n_{1},n_{2}\right)
}^{+};\quad V_{i}=\sum_{n_{1},n_{2}\in {\bf Z}}{\bf b}_{\left(
n_{1},n_{2}\right) }^{+}.  \label{12}
\end{equation}
Due to the remarkable properties ${\bf a}_{\left( n_{1},n_{2}\right)
}^{+}\pi _{\left( n_{1},n_{2}\right) }=\pi _{\left( n_{1}+1,n_{2}\right) }%
{\bf a}_{\left( n_{1},n_{2}\right) }^{+}$ , ${\bf b}_{\left(
n_{1},n_{2}\right) }^{+}\pi _{\left( n_{1},n_{2}\right) }=\pi _{\left(
n_{1},n_{2}+1\right) }{\bf b}_{\left( n_{1},n_{2}\right) }^{+}$\ and \ ${\bf %
b}_{\left( n_{1},n_{2}\right) }^{+}{\bf a}_{\left( n_{1},n_{2}\right)
}^{+}\pi _{\left( n_{1},n_{2}\right) }=\pi _{\left( n_{1}+1,n_{2}+1\right) }%
{\bf b}_{\left( n_{1},n_{2}\right) }^{+}{\bf a}_{\left( n_{1},n_{2}\right)
}^{+}$\ , it follows that $U$ and $V$ satisfy the following non commutative
algebra,
\begin{equation}
UV=e^{-i\psi \tau }VU,  \label{13}
\end{equation}
describing the complex extension of the CDS torus\cite{r1};\ to which we
shall refer to as the NC complex two torus. Since $\psi $\ is an arbitrary
complex parameter, eqs(\ref{13}) define an irrational NC complex two torus.

b) {\it Continuous Case}:\qquad In this case, the generator $T$ of ${\cal R}%
_{\left( con,con\right) }\left( C^{\ast }\right) $\ has a continuous
spectrum with a vector basis $\left\{ |z>,z\in {\bf C;\quad <}z{\bf ^{\prime
}|}z{\bf >=\delta }\left( z-z^{\prime }\right) \right\} $. It acts
diagonally as \ $<z|T|=z<z|$ and so $<z|U=\left( \exp i\psi z\right) <z|$
which imply in turns, $U=\int dz$ $\chi \left( \psi ,z\right) $ $\pi \left(
z\right) $ with $\chi \left( \psi ,z\right) =\exp \left( i\psi z\right) $
and $\pi \left( z\right) =|z><z|$; $\pi \left( z\right) \pi \left( z^{\prime
}\right) ={\bf \delta }\left( z-z^{\prime }\right) $ $\pi \left( z\right) $.
Since ${\cal R}_{\left( con,con\right) }\left( C^{\ast }\right) $ is
completely reducible, the $I_{id}$ identity operator may be decomposed as,
\begin{equation}
I_{id}=\int dz\text{ }\pi \left( z\right) .  \label{14}
\end{equation}
The shift operator by an $\epsilon $\ amount, denoted as $V\left( \epsilon
\right) $, acts on $\left\{ |z>,z\in {\bf C}\right\} $ as $<z|V\left(
\epsilon \right) =<z+\epsilon |$. It may be defined, by help of ${\bf a}%
^{+}\left( z,\epsilon \right) =|z><z+\epsilon |$ \ operators,\ as,
\begin{equation}
V\left( \epsilon \right) =\int dz\text{ }{\bf a}^{+}\left( z,\epsilon
\right) .  \label{15}
\end{equation}
These operators satisfy similar relations as in eqs(\ref{12},\ref{13})
namely $UV=\exp \left( -i\psi \tau \right) $ $VU$ and may also be realized
on the space of holomorphic functions $F\left( z\right) ={\bf <}z{\bf |}F%
{\bf >}$\ as
\begin{equation}
UFU^{-1}=\left( \exp i\psi z\right) F,\quad V_{\epsilon }FV_{\epsilon
}^{-1}=\left( \exp \epsilon \partial _{z}\right) F.  \label{16}
\end{equation}
We give details on the differential realization of $U$ and $V$ operators in
the next section.

2. {\it NC }${\bf C}^{\ast }${\it toric cycles}

\qquad If one forgets for a while about the quantum symmetry generated by $%
V_{a}$ and focus on the ${\bf C}^{\ast }$ toric generators $U_{a}$ only, one
may also build representations where the $r$ complex cycles of the ${\bf C}%
^{\ast r}$ group are non commuting. This is achieved by demanding $\left[
T_{a},T_{b}\right] =im_{\left[ ab\right] }\neq 0$ which is ensured by
introducing torsion among the ${\bf C}^{\ast }$ factors. Here also we should
distinguish between discrete and continuous spectrums. In the continuous
case for instance, the algebra of NC toric cycles is,
\begin{equation}
U_{a}U_{b}=\Lambda _{ab}^{m_{\left[ ab\right] }}\text{ }U_{b}U_{a};\quad
\Lambda _{ab}=\exp \left( -i\psi _{a}\psi _{b}\right) ,\quad \left[
T_{a},T_{b}\right] =im_{\left[ ab\right] }I_{id}.  \label{17}
\end{equation}
where $m_{ab}$ is a $r\times r$\ matrix. A realization of $T_{a}$ on the
space ${\cal F}$ of holomorphic functions $f\left( y_{1},...,y_{r}\right) $
with $r$ arguments, is given by,
\begin{equation}
\left[ T_{a},f\left( y_{1},...,y_{r}\right) \right] =\left( \partial
_{a}-im_{ac}y_{c}\right) f.  \label{18}
\end{equation}
Quantum symmetries may be also considered by allowing the \ $f\left(
y_{1},...,y_{r}\right) $\ \ functions to depend on extra variables $z_{a}$
so that we have new function $F\left( z_{a};y_{a}\right) $ satisfying,
\begin{equation}
U_{a}FU_{a}^{-1}=\left( \exp i\psi _{a}\left( z_{a}+\partial
_{y_{a}}-im_{ac}y_{c}\right) \right) \text{ }F,\quad V_{b}FV_{b}^{-1}=\left(
\exp \epsilon _{bd}\partial _{z_{d}}\right) \text{ }F  \label{19}
\end{equation}
\ Having studied the main lines of NC toric actions, we turn now to solve
the constraint eqs(\ref{8},\ref{9},\ref{10}).

\section{Representations of the $Z_{i}$'s}

The constraint eqs(\ref{8}) may be solved in different ways depending on
whether quantum symmetries are taken into account or not. In general,\ $\tau
_{ab}$\ torsion between $U_{a}$ and $V_{a}$ generators is introduced through
the relations $U_{a}V_{b}=\Omega _{ab}^{\tau _{ab}}V_{b}U_{a}$. The solution
for the $Z_{i}$'s reads as,
\begin{equation}
Z_{i}=x_{i}\prod_{a=1}^{r}\exp i\widetilde{q}_{i}^{a}\left( \psi
_{a}T_{a}+\phi _{a}S_{a}\right) ,  \label{20}
\end{equation}
where the $T_{a}$ and $S_{a}$ operators may be realized, for the
case of a continuous spectrum, as
\begin{equation}
T_{a}=\frac{\partial }{\partial y^{a}}+z_{a}-im_{ac}y_{c};\quad
S_{a}=\epsilon ^{ab}\frac{\partial }{\partial y^{b}}  \label{21}
\end{equation}
and where $x_{i}$\ are complex moduli, which we shall interpret as just the
commutative coordinates of the toric manifold ${\cal V}_{d+1}$.\ The $%
\widetilde{q}_{i}^{a}$s are shifted CY charges having extra contributions
coming from the toric data, $\widetilde{q}_{i}^{a}=q_{i}^{a}+%
\sum_{A=1}^{d}k_{A}$ $Q_{i}^{aA}+\sum_{\alpha }k^{\alpha }$ $N_{i\alpha
}^{a} $\ with $k_{A}$\ and $k^{\alpha }$ are numbers; they still obey $%
\sum_{i=0}^{k}\widetilde{q}_{i}^{a}=0$ which\ follow from the identities $%
\sum_{i=0}^{k}Q_{i}^{aA}=\sum_{i=0}^{k}N_{i\alpha }^{a}=0$ eqs (\ref{4}).
The $\theta _{ij}$\ parameters of the NC type IIA geometry we get are,
\begin{equation}
\theta _{ij}=\prod_{a,b=1}^{r}\Lambda _{ab}^{J_{ij}^{ab}}\Omega
_{ab}^{K_{ij}^{ab}},  \label{22}
\end{equation}
where now $J_{ij}^{ab}=m_{\left[ ab\right] }\widetilde{q}_{i}^{a}\widetilde{q%
}_{j}^{b}$ and $K_{ij}^{ab}\sim \tau _{\left[ ab\right] }\widetilde{q}%
_{i}^{a}\widetilde{q}_{j}^{b}$. By appropriate choices of $\Lambda _{ab}$, $%
\Omega _{ab}$\ , $m_{\left[ ab\right] }$ and $\tau _{\left[ ab\right] }$,
one recovers, as special cases, the\ representations involving discrete
torsions obtained in \cite{r15,r17}. According to the nature of spectrums of
$T_{a}$ and $S_{a}$, the $Z_{i}$ operators will have two sectors; discrete
and continuous. To illustrate the previous analysis, we consider the NC
extension of a CY manifold with a conic singularity. This manifold is
defined by a hypersurface ${\cal M}_{2}$ embedded in the toric variety $%
{\cal V}_{3}\subset {\bf C}^{6}{\bf /C}^{\ast 2}$ with ${\bf C}^{\ast 2}$
actions $x_{i}\longrightarrow x_{i}\exp i\left( \psi _{a}q_{i}^{a}\right) $
where $q_{i}^{1}=\left( 1,-1,1,-1,0,0\right) ,\quad q_{i}^{2}=\left(
1,0,-1,1,-1,0\right) $ and $p_{\alpha }=\left( 1,-1,1,-1\right) $. The $\nu
_{i}^{A}=\left( \nu _{i}^{1},\nu _{i}^{2},\nu _{i}^{3},\nu _{i}^{4}\right) $
and $\nu _{iA}^{\ast }=\left( \nu _{i1}^{\ast },\nu _{i2}^{\ast },\nu
_{i3}^{\ast },\nu _{i4}^{\ast }\right) $\ vertices satisfy $%
\sum_{i=1}^{6}q_{i}^{a}{\bf \nu }_{i}=0$ and $\sum_{\alpha =1}^{4}p_{\alpha }%
{\bf \nu }_{\alpha }^{\ast }=0$,
\begin{equation}
{\bf \nu }_{i}=\left(
\begin{array}{cccc}
1 & 1 & 0 & 0 \\
1 & 2 & 1 & 0 \\
1 & 1 & 2 & 1 \\
1 & 0 & 1 & 1 \\
1 & 0 & -1 & 0 \\
1 & 2 & 1 & 2
\end{array}
\right) ,\quad {\bf \nu }_{\alpha }^{\ast }=\left(
\begin{array}{cccc}
1 & 0 & 0 & 0 \\
1 & 1 & -1 & 0 \\
1 & 2 & -2 & 1 \\
1 & 1 & -1 & 1
\end{array}
\right) .  \label{23}
\end{equation}

The four $u_{\alpha }$ gauge invariants read as $u_{\alpha
}=\prod_{i=1}^{6}x_{i}^{n_{\alpha i}}$, $i=1,...,6$ \ with $n_{\alpha i}$\
integers as,
\begin{equation}
n_{\alpha i}=\left(
\begin{array}{cccccc}
1 & 1 & 1 & 1 & 1 & 1 \\
2 & 2 & 0 & 0 & 2 & 2 \\
3 & 3 & 0 & 0 & 3 & 5 \\
2 & 2 & 1 & 1 & 2 & 4
\end{array}
\right) .  \label{24}
\end{equation}
They satisfy the constraint eq $n_{0i}+n_{2i}=n_{1i}+n_{3i}$, showing that $%
{\cal V}_{3}$\ is described by $u_{0}u_{2}=u_{1}u_{3}$ with a conic
singularity at the origin. The complex two dimension CY hypersurface
embedded in ${\cal V}_{3}$, \ reads, in terms of the $x_{i}$ local
coordinates, as
\begin{equation}
P\left( x_{1},...,x_{6}\right)
=ax_{1}^{2}x_{2}^{2}x_{5}^{2}x_{6}^{2}+bx_{1}^{3}x_{2}^{3}x_{5}^{3}x_{6}^{5}+cx_{1}^{2}x_{2}^{2}x_{3}x_{4}x_{5}^{2}x_{6}^{4}+d\prod_{i=1}^{6}x_{i}
\label{25}
\end{equation}
The NC extension ${\cal M}_{2}^{nc}$ of this holomorphic hypersurface is
directly obtained. For the special case where the $L_{ij}$ antisymmetric
matrix is restricted to $L_{ij}=m\left(
q_{i}^{1}q_{j}^{2}-q_{j}^{1}q_{i}^{2}\right) $, with $L_{ij}=-L_{ji}$ and $%
L_{i6}=0$ and,
\begin{equation}
L_{ij}=m\left(
\begin{array}{ccccc}
0 & 1 & -2 & 2 & -1 \\
-1 & 0 & 1 & -1 & 1 \\
2 & -1 & 0 & 0 & -1 \\
-2 & 1 & 0 & 0 & 1 \\
1 & -1 & 1 & -1 & 0
\end{array}
\right) ,  \label{26}
\end{equation}
the NC complex surface ${\cal M}_{2}^{nc}$ is then given by a one parameter
algebra generated by following relations,
\begin{eqnarray}
Z_{1}Z_{2} &=&\Lambda ^{m}\text{ }Z_{2}Z_{1},\quad Z_{1}Z_{3}=\Lambda ^{-2m}%
\text{ }Z_{3}Z_{1},\qquad Z_{1}Z_{4}=\Lambda ^{2m}\text{ }Z_{4}Z_{1},
\nonumber \\
Z_{1}Z_{5} &=&\Lambda ^{-m}\text{ }Z_{5}Z_{1},\qquad Z_{2}Z_{3}=\Lambda ^{m}%
\text{ }Z_{3}Z_{2},\qquad Z_{2}Z_{4}=\Lambda ^{-m}\text{ }Z_{4}Z_{2},
\nonumber \\
Z_{2}Z_{5} &=&\Lambda ^{m}\text{ }Z_{5}Z_{2},\qquad Z_{3}Z_{4}=\text{ }%
Z_{4}Z_{3},\qquad Z_{3}Z_{5}=\Lambda ^{-m}\text{ }Z_{5}Z_{3},  \label{27} \\
Z_{4}Z_{5} &=&\Lambda ^{m}\text{ }Z_{5}Z_{4},\qquad Z_{i}Z_{6}=Z_{6}Z_{i},
\nonumber
\end{eqnarray}
where $\Lambda ^{m}$\ is given by $\Lambda ^{m}=\exp \left( -im\psi _{1}\psi
_{2}\right) $. Since $\psi _{a}=\rho _{a}-i\alpha _{a}$; it follows that $%
m\psi _{1}\psi _{2}=m\left( \rho _{1}\rho _{2}-\alpha _{1}\alpha _{2}\right)
-im\left( \alpha _{1}\rho _{2}+\alpha _{2}\rho _{1}\right) $ which we set as
$\Lambda ^{m}=\exp \left( \kappa +i\phi \right) $\ for simplicity. This is a
one complex parameter algebra enclosing various special situations
corresponding to: (1) \ {\it Hyperbolic representation} described by $\left(
\kappa ,\phi \right) \equiv \left( \kappa ,0\right) $. (2) {\it Periodic
representations} corresponding to $\left( \kappa ,\phi \right) \equiv \left(
0,\phi +2\pi \right) $ where $\left| \Lambda ^{m}\right| =1$. (3) {\it %
Discrete periodic representations} $\left( \kappa ,\phi \right) \equiv
\left( 0,N\phi +2\pi \right) $ with $\left| \Lambda ^{m}\right| =1$ but
moreover $\left( \Lambda ^{m}\right) ^{N}=1$. This last case is naturally a
subspace of the periodic representation and it is precisely the kind of
situation that happens in the building of BL NC manifolds with discrete
torsion.

\section{Fractional Branes}

The NC type $IIA$ realization we have studied concerns regular points of the
algebra. In this section, we want to complete this analysis by considering
the representations at fixed points where live fractional $D$ branes. To do
so, we shall first classify the various subspaces ${\cal S}_{(\mu )}$ of
stable points under ${\bf C}^{\ast r}$; then we give the quiver diagrams
extending those of Berenstein and Leigh\cite{r24}.

\begin{itemize}
\item  {\sl Fixed points}
\end{itemize}

The local holomorphic coordinates $\left\{ x_{i}\in {\bf C}^{k+1};\quad
0\leq i\leq k\right\} $ eq(1) are not all of them independent as they are
related by the ${\bf C}^{\ast r}$ gauge transformations $U_{a}:x_{i}%
\longrightarrow U_{a}x_{i}U_{a}^{-1}=x_{i}\lambda ^{q_{i}^{a}}$, with $%
\sum_{i=0}^{k}q_{i}^{a}=0$.\ Fixed points of the ${\bf C}^{\ast r}$ gauge
transformations are given by the solutions of the constraint eq
\begin{equation}
x_{i}=U_{a}x_{i}U_{a}^{-1}.  \label{28}
\end{equation}
Solutions\ of this eq depend on the values of $q_{i}^{a}$; the $x_{i}$'s
should be zero unless $q_{i}^{a}=0$. Fixed points are then given by the $%
{\cal S}$ subspace of ${\bf C}^{k+1}$ whose $x_{i}$ local coordinates are $%
{\bf C}^{\ast r}$\ gauge invariants. To get a more insight into this
subspace it is interesting to note that as ${\bf C}^{k+1}{\bf /C}^{\ast
r}=\left( {\bf C}^{k+1}{\bf /C}^{\ast }\right) {\bf /C}^{\ast r-1}$; it is
useful to introduce the ${\cal S}_{(a)}=\left\{ x_{i}\text{ \ }|\text{ \ }%
q_{i}^{a}=0;\qquad i\in J\subset \left\{ 0,1,...,k\right\} \right\} $\
subspaces that are invariant under the $a-th$ factor of the ${\bf C}^{\ast
r} $ group. So the manifold ${\cal S}$ stable under the full ${\bf C}^{\ast
r}$ is given by the intersection of the various ${\cal S}_{(a)}$'s, i.e

\begin{equation}
{\cal S}=\cap _{a=1}^{r}{\cal S}_{(a)}  \label{29}
\end{equation}
If we denote by $\left\{ x_{i_{0}};...;x_{i_{k_{0}-1}}\right\} $ those local
coordinates that have non zero $q_{i}^{a}$ charges; $J=\left\{
i_{0},...,i_{k_{0}}\right\} $, and $\left\{
x_{i_{k_{0}-1}};...;x_{i_{k}}\right\} $\ the coordinates that are fixed
under ${\bf C}^{\ast r}$; then the manifold ${\cal S}$ is given by,
\begin{equation}
{\cal S}=\left\{ \left( 0,...,0,x_{k_{0}},...,x_{k}\right) \right\} \subset
{\cal V}_{d+1}\subset {\bf C}^{k+1}  \label{30}
\end{equation}
To get the representation of the $Z_{i}$ variable operators on ${\cal S}$,
let us first consider what happens on its neighboring space ${\cal S}%
^{\epsilon }=\left\{ \left( \epsilon ,...,\epsilon ,x_{k_{0}}+\epsilon
,...,x_{k}+\epsilon \right) \right\} $, where $\epsilon $\ is as small as
possible. Using the hypothesis $q_{i_{k_{0}}}^{a}=...=q_{i_{k}}^{a}=0$ \ we
have made and replacing the $x_{i}$ moduli by their expression on ${\cal S}%
^{\epsilon }$, then putting in the realization eqs(20), we get the following
result,
\begin{eqnarray}
Z_{i_{j}} &=&\epsilon \prod_{a=1}^{r}\exp iq_{i}^{a}\left( \psi
_{a}T_{a}+\phi _{a}S_{a}\right) ;\quad 0\leq j\leq k_{0}-1,  \label{31} \\
Z_{i_{j}} &=&\left( x_{i_{j}}+\epsilon \right) I_{id};\quad k_{0}\leq j\leq
k.  \label{32}
\end{eqnarray}
The representation of the $Z_{i}$'s\ on ${\cal S}$\ is then obtained by
taking the limit $\epsilon \longrightarrow 0$.\ As such non trivial
operators are given by $Z_{i_{j}}\sim x_{i_{j}}I_{id};\quad k_{0}\leq j\leq
k $; they are proportional to the $I_{id}$ operator of ${\cal R}\left(
C^{\ast r}\right) $. This an important point since the $Z_{i}$\ operators
are reducible into an infinite component sum as shown here below,\footnote{%
The sums involved in the decomposition of the identity are either discrete
series, integrals or both of them depending on whether the group
representation ${\cal R}\left( C^{\ast r}\right) $ spectrum is discrete,
continuous or with discrete and continuous sectors. Therefore we have either
$I_{id}=\sum_{{\bf n}}\pi _{{\bf n}}$, $\ {\bf n=}\left(
n_{1},...,n_{r}\right) $; $\ I_{id}=\int d{\bf \sigma }\pi \left( {\bf %
\sigma }\right) $, ${\bf \sigma =}\left( \sigma _{1},...,\sigma _{r}\right) $%
; or again $\sum_{{\bf m}}\int d{\bf \zeta }\pi _{{\bf m}}\left( {\bf \zeta }%
\right) $; Here $\ \pi _{{\bf m}}$, $\ {\bf m=}\left(
n_{i_{1}},...,n_{i_{r_{0}}}\right) $ \ and $\pi \left( {\bf \zeta }\right) $%
, \ ${\bf \zeta =}\left( \zeta _{i_{r_{0}+1}},...,\zeta _{i_{r}}\right) $; \
and are the $C^{\ast r}$ representation projectors considered in section 3 .}
\begin{equation}
Z_{i}=\sum_{{\bf n}}\text{ }Z_{i}^{\left( {\bf n}\right) };\quad
Z_{i}^{\left( {\bf n}\right) }=x_{i}\pi _{{\bf n}}.  \label{33}
\end{equation}

\begin{itemize}
\item  {\sl Brane interpretation}
\end{itemize}

The above decomposition of the $Z_{i}$'s on ${\cal S}$ has a nice
interpretation in the $D$ brane language. Thinking about $x_{i}$'s as the
coordinates of a $D$ $p$ brane, ( $p=2d$), wrapping $M_{d}$, it follows
that, due NC torus fibration, $D$ $p$ at singular points fractionate. In
addition to the results of \cite{r24}, which apply as well for the present
study, there is a novelty here due to the dimension of the completely
reducible ${\cal R}\left( C^{\ast r}\right) $ representation. There are
infinitely many values for the $C^{\ast }$ characters and so an infinite
number of fractional $D\left( p-2k_{0}\right) $ branes wrapping ${\cal S}$.
However, this is a unacceptable solution from string compactification point
of view. Branes should have finite tensions in string theory background
unless there are non compact dimensions. In fact this is exactly what we
have since the irrational representation of the NC torus fibration
introduces extra non compact directions as shown on the realization eqs(\ref
{20},\ref{21}). Such behaviour has no analogue in the BL case.

\begin{itemize}
\item  {\sl Quiver Diagrams}
\end{itemize}

Like for the BL case of CY orbifolds, one can here also describe the
varieties of fractional $D$ branes by generalizing the BL quiver diagrams
for ${\cal M}_{d}^{nc}$ at fixed points. One of the basic ingredients in
getting these graphs is the identification of the projectors of ${\bf C}%
^{\ast r}$ and the step operators ${\bf a}^{\pm }$. Since each ${\bf C}%
^{\ast }$ factor has completely reducible representations with four sectors,
it is interesting to treat them separately. The various sectors for each $%
{\bf C}^{\ast }$ subsymmetry factor are as follows; {\it (i)} $\left(
dis,dis\right) $ discrete-discrete sector where the $C^{\ast }$ characters
are given by $\chi _{n}\left( \psi \right) =\exp i\psi n$; $n\in {\bf Z+}i%
{\bf Z,}$ $\psi \in {\bf C}$; {\it (ii)} $\left( dis,con\right) $
discrete-continuous and \ $\left( con,dis\right) $ \ continuous-discrete
sectors \ and finally {\it (iii)} \ $\left( con,con\right) $
continuous-continuous sector with characters $\chi \left( p,\psi \right)
=\exp ip\psi $; $p\in {\bf C}$. Recall that, due to torsion, the algebraic
structure of the $D$ $p$ branes wrapping the NC manifold change. Brane
points $\{x_{i}\}$ of commutative type $IIA$ geometry become fibers based on
$\{x_{i}\}$. These fibers are valued in the algebra of the group
representation ${\cal R}\left( C^{\ast r}\right) $ and may be given a simple
graph description. While points $x_{i}.1$ in the commutative type $IIA$
geometry are essentially numbers, the $Z_{i}$ coordinate operators can be
thought of as
\begin{equation}
x_{i}.1=\ \rightarrow \ \ Z_{i}=\left( Z_{i}\right) _{mn}\ U^{m}V^{n}.
\label{34}
\end{equation}
Extending the results of \cite{r24}, one can draw graphs for fractional $D$
branes. Due to the decomposition of $I_{id}$ eqs(\cite{r11}) and (\ref{14}),
we associate to each $D$ $p$ brane coordinate a quiver diagram mainly given
by the product of ( discrete or continuous )\ $S^{1}$\ circles. For the
simplest case $r=1$ and $C^{\ast }$ discrete representations, the quiver
diagram is built as follows: {\it (1)} To each $\pi _{n}$ projector it is
associated a vertex point on a discrete $S^{1}$ circle. As there is an
infinite number of points that one should put on $S^{1}$, all happens as if
the quiver diagram is given by the ${\bf Z+}i{\bf Z}$ \ lattice plus a extra
point at infinity. {\it (2)} The $a_{n}^{\pm }$ shift operators are
associated with the oriented links joining adjacent vertices, ( vertex $%
\left( n-1\right) $ to the vertex $\left( n\right) $ for $a_{n}^{-}$\ and
vertex $\left( n\right) $ to the vertex $\left( n+1\right) $ for $a_{n}^{+}$%
\ ) of quiver diagram. They act as automorphisms exchanging the ${\bf C}%
^{\ast }$\ characters.

Moreover, as $D$ $p$ brane coordinates at the singularities are of the form $%
Z_{i}\sim \sum_{n}$ $Z_{i}^{\left( n\right) }$, it follows that $D$ $p$
branes on ${\cal S}$ sub-manifolds fractionate into an infinite set of
fractional $D2s$ branes coordinated by $Z_{i}^{\left( n\right) }$. This is a
remarkable feature which looks like the process of tachyon condensation
mechanism \`{a} la GMS \cite{r8,r10,r11,r12} where for instance a non
compact $D$ $p_{1}$ brane on a NC Moyal plane decomposes into an infinite
set of non compact $D$ $\left( p_{1}-2\right) $ branes. Such discussion is
also valid for the case of a continuous spectrum; the corresponding quiver
diagram is given by cross products of circles.

\section{Conclusion}

Using the BL algebraic geometry approach, we have studied\ the type $IIA$
geometry of NCCY manifolds embedded in toric varieties ${\cal V}$ endowed
with a NC toric fibration. Actually this study completes partial results of
works in the literature on NCCY manifolds and too particularly orbifolds of
CY homogeneous hypersurfaces with discrete torsion. The results of this
paper concerns a more general class of singular CY manifolds ${\cal M}_{d}$
\ embedded in toric varieties ${\cal V}_{d+1}$ with ${\bf C}^{\ast r}$ toric
actions endowed by torsions. These torsions are carried either by quantum
symmetries described by inner automorphisms of ${\bf C}^{\ast r}$ or again
by considering NC complex cycles within the toric group factors in the same
manner as one does in the Connes et {\it al} solution for toroidal
compactification of matrix model of M theory. Among our results, we find
that complex $d$ dimension\ NCCY manifolds ${\cal M}_{d}^{nc}$ are non
compact manifold that are naturally described in the language of toric
geometry. The $\theta _{ij}$ parameters carrying the structure have
contributions involving, in addition to the usual CY condition $%
\sum_{i=0}^{k}q_{i}^{a}=0$, the data of the toric polygons using the
relations $\sum_{i=0}^{k}q_{i}^{a}\nu _{i}^{A}=0$.


\begin{thebibliography}{99}
\bibitem{r1}  A. Connes, M.R. Douglas et A. Schwarz,JHEP 9802, 003 (1998).

\bibitem{r2}  David J. Gross, Nikita A. Nekrasov, 0103 (2001) 044..

\bibitem{r3}  N. Seiberg and E. Witten, JHEP 9909(1999) 032..

\bibitem{r4}  N. Nekrasov and A. Schwarz, Commun Math. Phys 198(1998)
689-703..

\bibitem{r5}  A. Belhaj, M. Hssaini, E. Sahraoui, E. H. Saidi, CQG. 18(2001)
2339.

\bibitem{r6}  Ofer Aharony, Micha Berkooz, JHEP 9910 (1999)030.

\bibitem{r7}  A. Belhaj, E.H.Saidi, Mod.Phys.Lett A15 (2000) 1767, CQG18
(2001) 57.

\bibitem{r8}  R. Gopakumar, S. Minwalla and A. Stominger, JHEP 0005
(2000)020.

\bibitem{r9}  E. M. Sahraoui, E.H Saidi, hep-th/0012259, Class.Quant.Grav.
18 (2001) 3339-3358.

\bibitem{r10}  I Bars, H Kajiura, Y Matsuo, T Takayanagi, Phys.RevD63 (2001)
086001.

\bibitem{r11}  B. R. Greene, C. I. Lazaroiu, Piljin Yi, Nucl.Phys. B539
(1999) 135-165.

\bibitem{r12}  E. M. Sahraoui, E.H. Saidi, JHEP 0205 (2002) 063,
hep-th/0105188.

\bibitem{r13}  I. Benkaddour , M. Bennai, E. Diaf and H. Saidi CQG
17(2000)1765.

\bibitem{r14}  David Berenstein, Vishnu Jejjala, Robert G. Leigh, Nucl.Phys.
B589 (2001).

\bibitem{r15}  David Berenstein, Robert G. Leigh, Phys.Lett. B499 (2001)
207-214.

\bibitem{r16}  Hoil Kim, Chang-Yeong Lee, hep-th/0105265;

\bibitem{r17}  A. Belhaj and E. H, Saidi, Lab/UFR-HEP 0110.

\bibitem{r18}  E.H Saidi, hep-th/0202104

\bibitem{r19}  S. Katz, P. Mayr and C. Vafa, Adv, Theor. Math. Phys
1(1998)53.

\bibitem{r20}  A. Belhaj, A. E. Fallah , E. H, Saidi, CQG 17 (2000)515-532.

\bibitem{r21}  M. Cvetic, G.W. Gibbons, H. Lu, C.N. Pope, Nucl.Phys. B606
(2001) 18.

\bibitem{r22}  Subir Mukhopadhyay, Koushik Ray, JHEP 0107 (2001) 007.

\bibitem{r23}  M. Petrini, R. Russo, A. Zaffaroni, Nucl.Phys. B608 (2001)
145.

\bibitem{r24}  David Berenstein, Robert G. Leigh, JHEP 0106 (2001) 030.

\bibitem{r25}  N.C. Leung and C. Vafa; Adv .Theo . Math. Phys 2(1998) 91.

\bibitem{r26}  S.Katz, P. Mayr, C. Vafa, Adv.Theor.Math.Phys. 1 (1998) 53.

\bibitem{r27}  A. Belhaj, E.H Saidi, hep-th/0012131.

\bibitem{r28}  P Candelas, E Perevalov, G Rajesh, Nucl. Phys. B450 (1995)
267.

\bibitem{r29}  Wolfgang Lerche, Cumrun Vafa, Nicholas P. Warner;
Nucl.Phys.B324:427,1989
\end{thebibliography}
\end{document}